# Phase separation in mixed argon-xenon clusters

O.G. Danylchenko[1], Yu.S. Doronin, S.I. Kovalenko, V.N. Samovarov

*B.I. Verkin Institute for Low Temperature Physics and Engineering
of the National Academy of Sciences of Ukraine,
47 Lenin Ave., 61103, Kharkiv, Ukraine*

Abstract

*In this paper, electron diffraction and optical methods are used together for the first time to study the problem of phase equilibrium in binary clusters of heavy rare gases. For the argon-xenon system, a new effect of its decay into pure components with a sharp interface between the xenon core and the argon shell is observed.*

Solid solution phase diagrams of bulk rare gas cryocrystals have been studied rather in detail. For the case of heavy rare gases such as argon, krypton, and xenon, the formation of binary solid solutions with both finite and infinite solubility takes place. Some peculiarities of phase equilibrium in two-component rare gas clusters are still to be studied. There is some theoretical [1] and experimental [2] evidence that two-component rare gas clusters may exhibit radial segregation of atoms.

In this paper, electron diffraction and optical methods are used together for the first time to study the problem of phase equilibrium in binary clusters of heavy rare gases. Substrate-free mixed clusters were formed in a supersonic gas jet expanding into vacuum [3]. To change cluster size, we varied the pressure $P_0$ (0.25 – 2.5 bar) and the temperature $T_0$ (250 – 100 K) of the gas at the entrance to the conical nozzle (340 μm, $2\Theta = 8.6^0$). Xenon impurity concentration was varied in original gas mixtures from $1\times10^{-3}$ through 3 at. %. The cluster beam was studied ≈ 30 mm downstream from the nozzle, cluster temperature being there ≈ 40 K. For comparison reasons, we performed analogous measurements on argon-krypton gas mixtures. Both crystalline (fcc structure) and quasicrystalline (icosahedral structure) clusters were formed in the experiments, the cluster size being determined from electron diffraction data [4].

---
[1] e-mail: danylchenko@ilt.kharkov.ua

Fig. 1 shows diffraction patterns from the clusters formed in pure argon (1a) and in argon with 2.5 at. % of krypton (1b); fig. 2 shows analogous results for the argon-xenon mixture with 0.5 at. % of xenon. As we can see from fig. 1, there is a shift $\Delta s$ of the diffraction maxima for the argon-krypton clusters, $\Delta s/s_{Ar} = (2 \pm 0.1)\%$ (the (220) peak shift is marked with dashed lines as an example). The shift indicates unambiguously the formation of a substitutional solution with 25 at. % of krypton in a cluster which is one order of magnitude more than in the original gas mixture[1]. A quite different picture is observed for the case of argon-xenon. As is seen from fig. 2, the argon peaks are accompanied by xenon peaks, the (111) xenon maximum being most pronounced. This result is indicative of nonsolubility of the components and of the presence of two aggregations in a cluster, one of which is composed of argon atoms, while the other contains mainly xenon.

Optical cathodoluminescence spectra of the clusters not only confirm the decay of the argon-xenon system but also provide evidence of phase separation into pure components. Fig. 3 shows VUV luminescence spectra for the cases of 0.005 and 0.5 at. % of xenon in argon. Fig. 4 demonstrates quenching of excited molecules which are characteristic of argon-xenon solution: $(Ar_2^*)_{Xe}$ excimer molecules (excited argon centers in xenon matrix) and $(ArXe)^*$ exciplex molecules. From fig. 3 and 4 it can be seen that for xenon concentrations 0.5 at. % and more the spectra contain only luminescence of the xenon core, which "intercepts" the electronic excitation originating in the argon shell. It should be noted that for argon-krypton solutions in clusters, the luminescence of the $(ArKr)^*$ excimer molecules is observed over the whole interval of krypton impurity concentration ($1\times10^{-3}$ – 3 at. %) [6].

---

[1] The enrichment effect for mixed clusters is discussed in more detail in Ref. [5].

Cathodoluminescence spectra of rare gases in the VUV region make it possible to determine extremely small quantities (~ $10^{-3} - 10^{-4}$ at. %) of impurity atoms embedded in a matrix. Estimations show that we could easily detect the presence of single xenon atoms in the argon shell of clusters (each containing several hundreds of atoms, overall number of clusters being $10^{13}$-$10^{14}$). Thus, the optical data are indicative of a decay of the argon-xenon clusters into practically pure components. For bulk argon-xenon systems at T < 70 K, only a decay into two solutions with the concentration of one component in the other ≈ 10 at. % was observed [7].

Thus, we show experimentally that in mixed argon-xenon solutions a decay of the system into pure components with the formation of a sharp interface between the xenon core and the argon shell takes place. For bulk argon-xenon samples, the present phase diagram excludes such a segregation. The observed phenomenon may be due to some nanoscale properties of the argon-xenon clusters.

**Captions**

Fig. 1. Diffraction patterns from substrate-free clusters: (a) – pure argon; (b) – argon with 2.5 at. % of krypton. The shift Δs indicates the formation of a substitutional solution.

Fig. 2. Diffraction patterns from substrate-free clusters: (a) – pure argon; (b) – argon with 0.5 at. % of xenon. The fcc structure peaks are marked with dot-and-dash lines for argon ($a_{Ar}$=5.34 Å) and xenon ($a_{Xe}$=6.146 Å).

Fig. 3. Cathodoluminescence spectra of substrate-free mixed argon-xenon clusters: (a) – 0.005 at. % of xenon, the inset shows a part of the spectrum near 7.7 eV at a larger scale; (b) – 0.5 at. % of xenon.

Fig. 4. Quenching curves of argon and argon-xenon molecular centers in mixed clusters as a function of xenon concentration in the original gas mixture: (a) – $Ar_2^*$ and $(Ar_4^+)^*$ centers; (b) – $(ArXe)^*$ and $(Ar_2^*)_{Xe}$ centers.

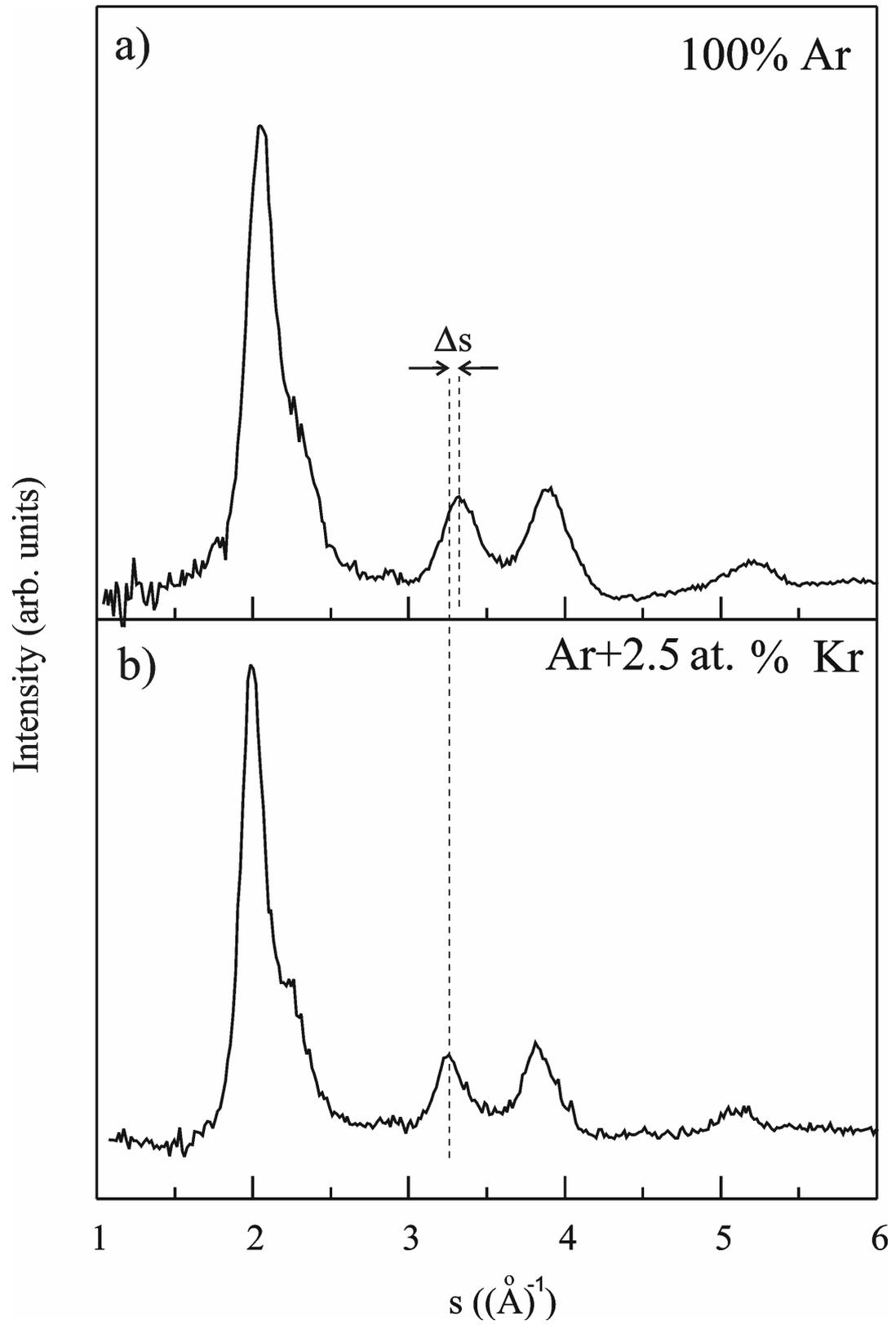

Fig. 1.

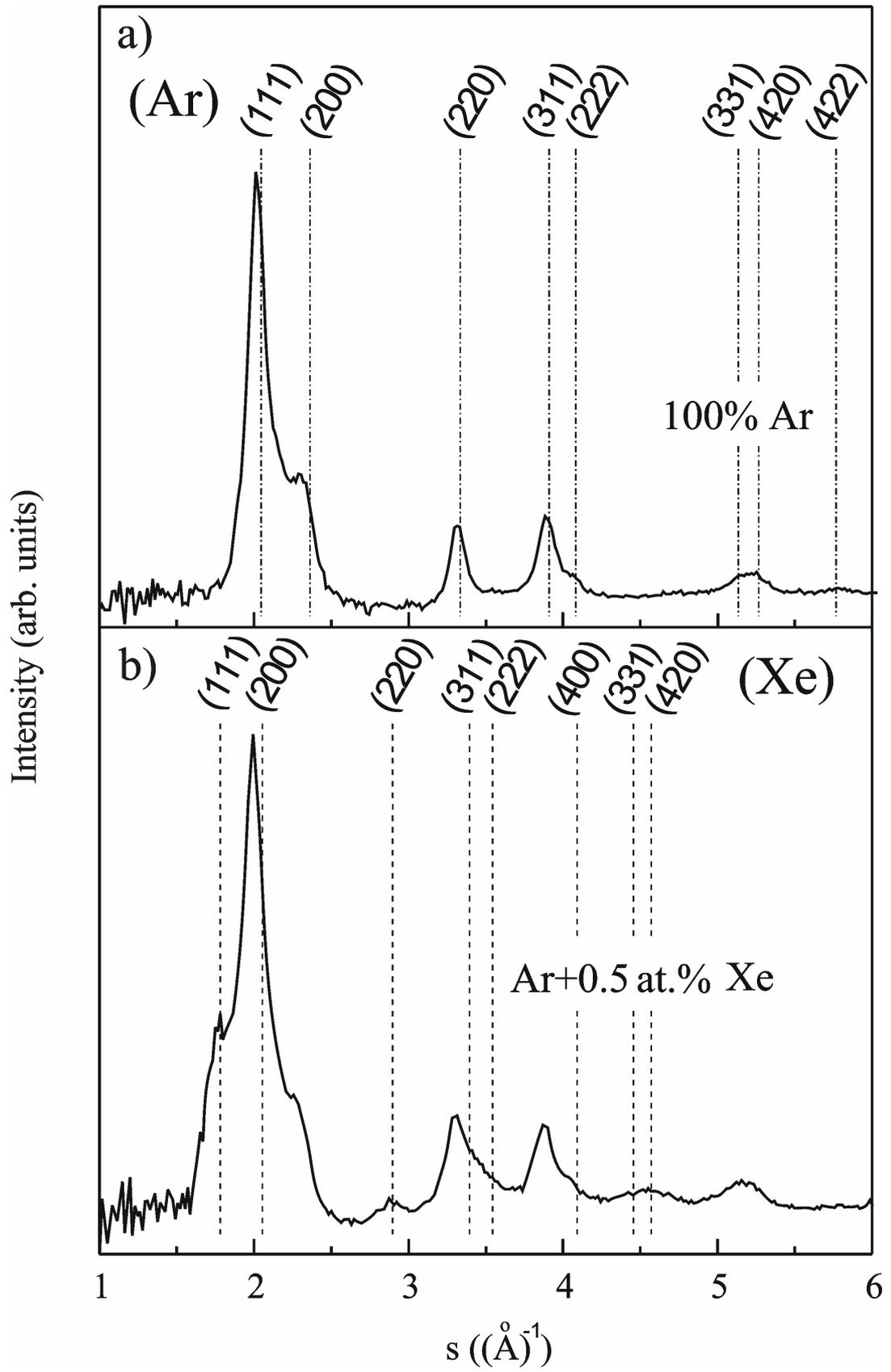

Fig. 2.

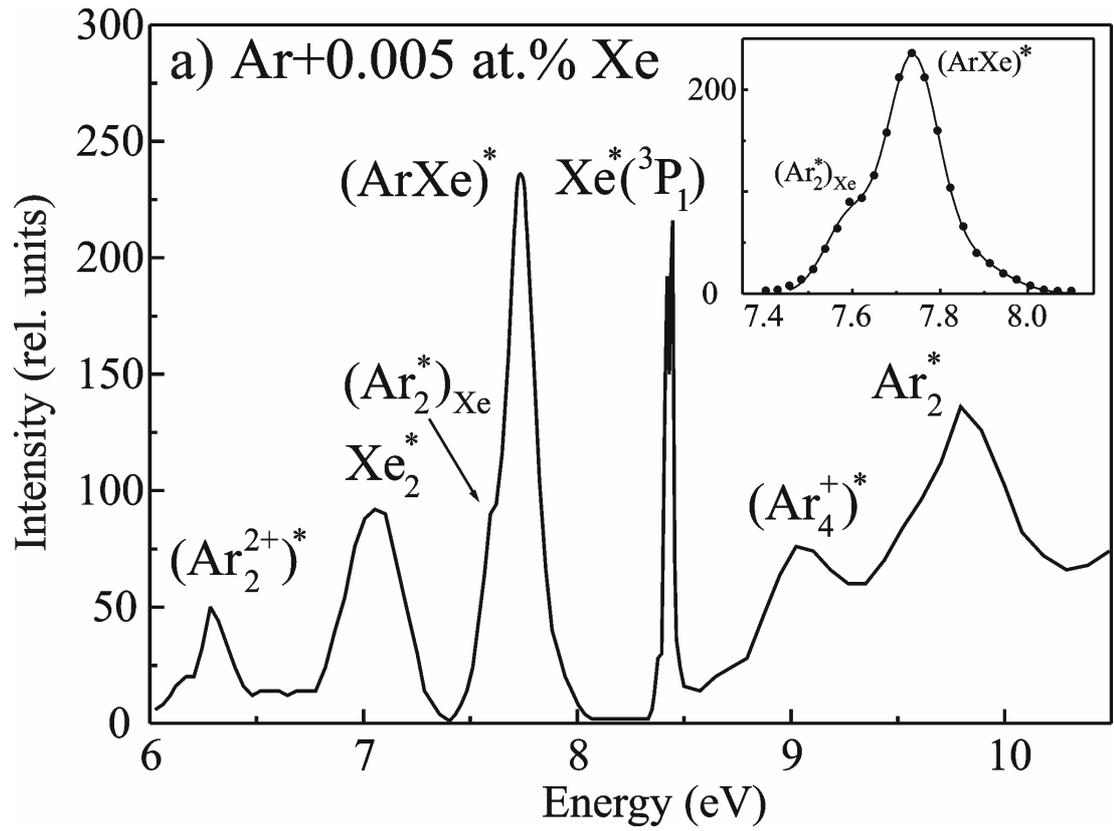
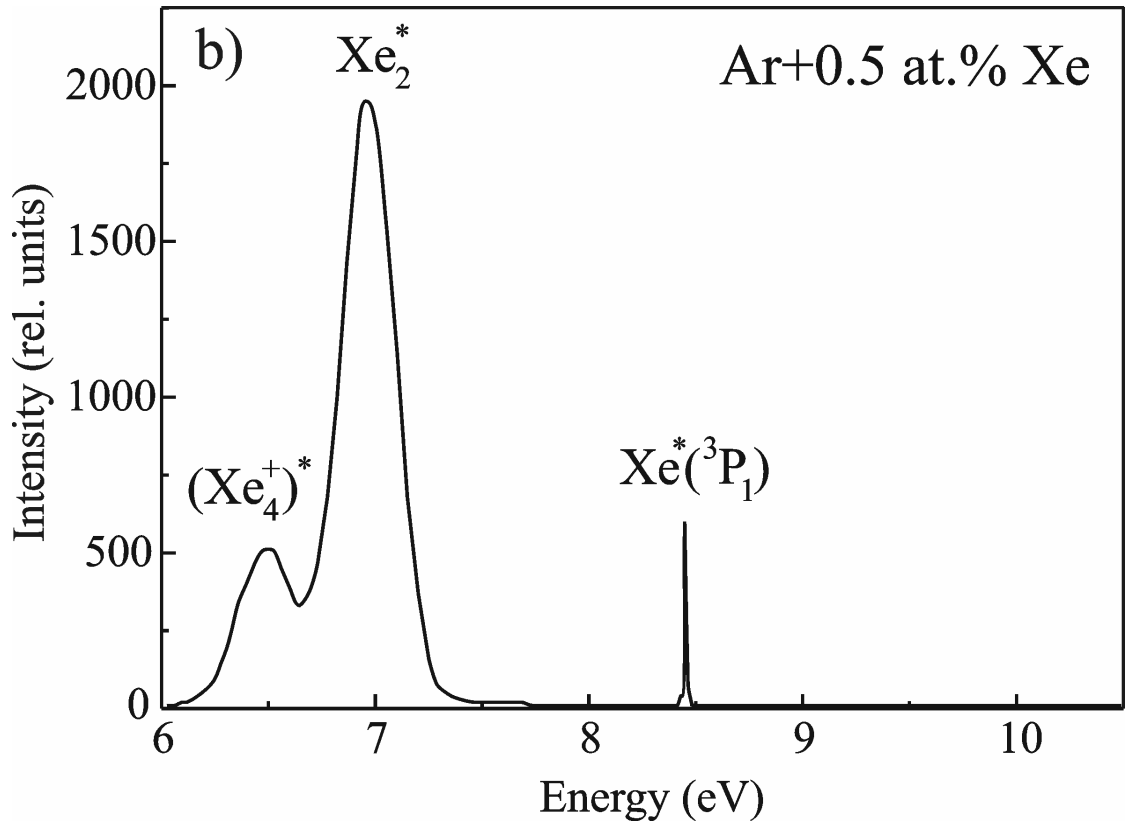

Fig. 3.

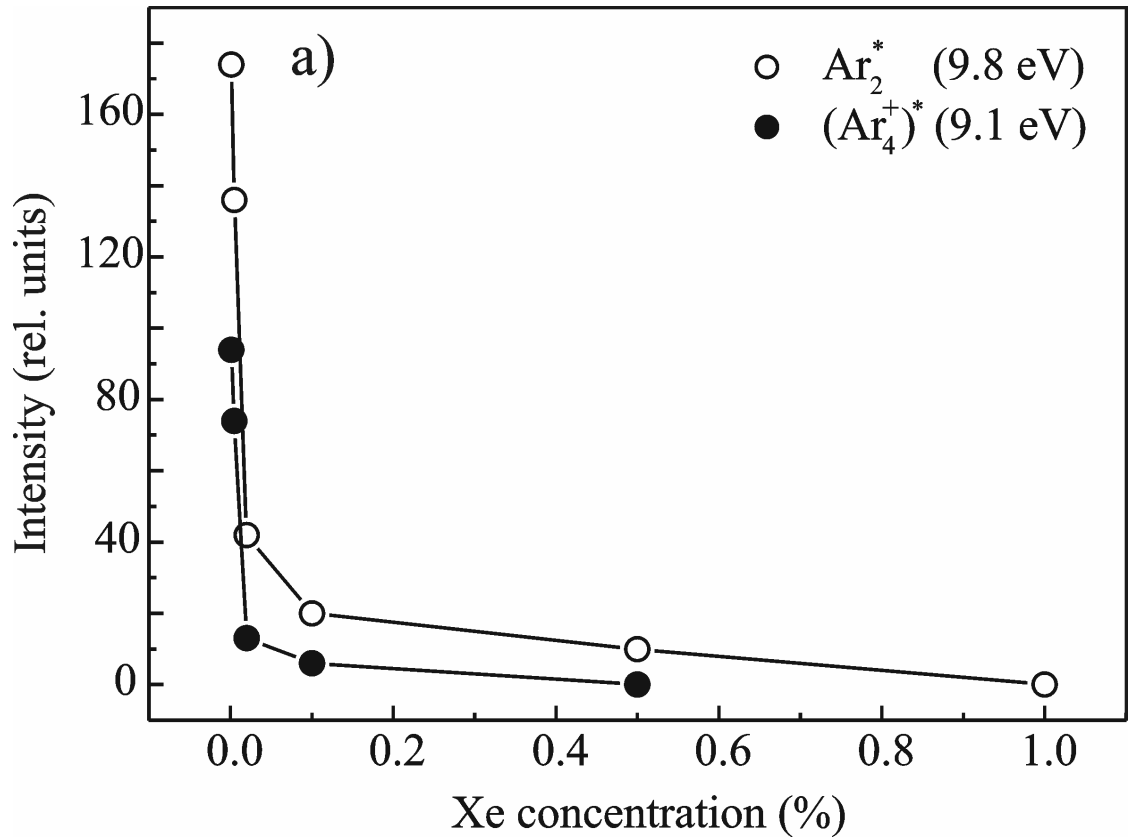
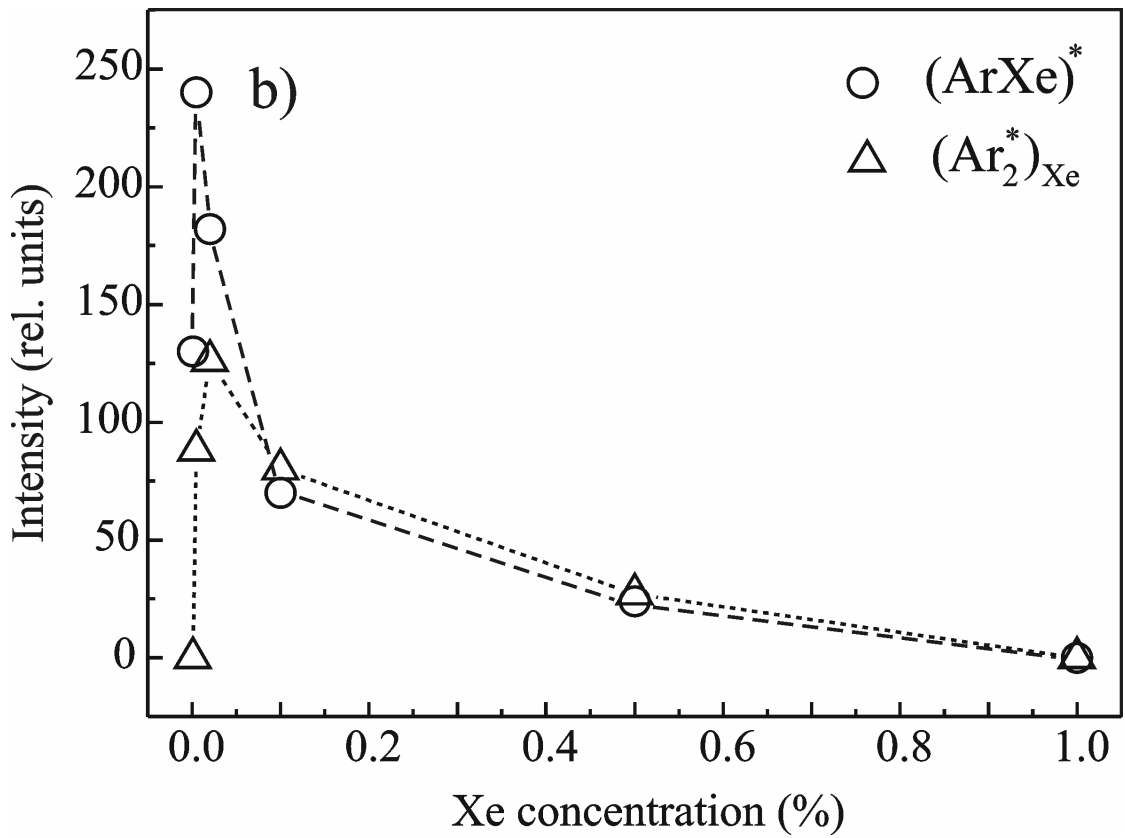

Fig. 4.